\documentclass[prb,reprint,twocolumn,floatfix,twoside,tbtags,showpacs]{revtex4-1}

\usepackage[utf8]{inputenc}
\usepackage{graphicx}
\usepackage{subfigure}
\usepackage{amsmath,amssymb}
\usepackage{bm}	
\usepackage{xspace} 
\usepackage[]{hyperref}

\DeclareMathOperator{\Max}{Max}

\newcommand{\etal}{\textit{et al}.\@\xspace}
\newcommand{\ie}{\textit{i.e.}\@\xspace}
\newcommand{\cf}{\textit{cf}.\@\xspace}
\newcommand{\siesta}{\textsc{Siesta}\@\xspace}
\newcommand{\pwscf}{\textsc{Pwscf}\@\xspace}
\newcommand{\vasp}{\textsc{Vasp}\@\xspace}
\newcommand{\abinitio}{\textit{ab initio}\@\xspace}
\newcommand{\Abinitio}{\textit{Ab initio}\@\xspace}

\begin{document}

\title{Screw dislocation in zirconium: an \abinitio study}

\author{Emmanuel \surname{Clouet}}
\email{emmanuel.clouet@cea.fr}
\affiliation{CEA, DEN, Service de Recherches de Métallurgie Physique,
91191 Gif-sur-Yvette, France}


\date{\today}
\begin{abstract}
  Plasticity in zirconium is controlled by $1/3\langle1\bar{2}10\rangle$ screw dislocations
  gliding in the prism planes of the hexagonal close-packed structure. 
  This prismatic and not basal glide is observed for a given set of transition metals like zirconium
  and is known to be related to the number of valence electrons in the d band.
  We use \abinitio calculations based on the density functional theory to study the core structure
  of screw dislocations in zirconium. 
  Dislocations are found to dissociate in the prism plane in two partial dislocations,
  each with a pure screw character.
  \Abinitio calculations also show that the dissociation in the basal plane is unstable.
  We calculate then the Peierls barrier for a screw dislocation gliding in the prism plane
  and obtain a small barrier.
  The Peierls stress deduced from this barrier is lower than 21\,MPa, 
  which is in agreement with experimental data.
  The ability of an empirical potential relying on the embedded atom method (EAM)
  to model dislocations in zirconium is also tested against these \abinitio calculations.
\end{abstract}
\maketitle

\section{Introduction}

Plasticity in $\alpha$-zirconium is controlled by 
dislocations with a $1/3\langle1\bar{2}10\rangle$ Burgers vector
gliding in the prism planes of the hexagonal compact (hcp) lattice.
\cite{Rapperport1959,Soo1968,Akhtar1971,Caillard2003}
The relative ease of prismatic glide compared to basal glide
has been shown to be linked to the 
ratio of the corresponding stacking fault energies, 
which in turn is controlled by the electronic structure. \cite{Legrand1984}
In particular, Legrand\cite{Legrand1984} used a tight binding model 
to show that prismatic slip in transition metals of the IV\,B column (Zr, Ti, Hf)
originates from the electronic filling of the valence d band.
As a consequence, it appears necessary to take into account 
the anisotropy of the d orbital, and hence the angular dependence 
of the atomic bonding, to model dislocations in these transition
metals \cite{Legrand1986}.
One cannot therefore rely on central forces empirical potential
and needs to take account of the electronic structure.
Tight binding models \cite{Legrand1985,Girshick1998b}
or \abinitio calculations 
\cite{Ferrer2002,Domain2001,Domain2006,Tarrat2009,Ghazisaeidi2012}
show indeed that a $1/3\langle1\bar{2}10\rangle$ screw dislocation, either in Zr or in Ti, 
spreads in prism planes, in agreement with the prismatic glide 
observed experimentally.
But none of these atomistic simulations
calculate the Peierls stress of a screw dislocation.
According to some authors,\cite{Legrand1985,Vitek2008} its core structure 
is not completely planar, which may be 
the cause of a high Peierls stress.

It is true, experimentally, that screw dislocations 
glide with difficulty compared to other dislocation characters
in zirconium or titanium alloys: 
characteristic microstructures, with long and straight dislocations
aligned along their screw orientation, 
are observed at low temperature,
\cite{Bailey1962,Akhtar1971,Ferrer2002,Naka1988,Farenc1993,Farenc1995}
and the flow stress is strongly temperature dependent, 
\cite{Mills1968,Soo1968,Akhtar1971,Sastry1971,Sastry1972,Tanaka1972,Akhtar1975a,Naka1988}
in agreement with the assumption of a high Peierls barrier 
which must be overcome by the nucleation of double kinks.
But experiments also show that the yield stress in zirconium or in titanium
strongly decreases with a decreasing amount of interstitial impurities
like oxygen.
\cite{Mills1968,Soo1968,Akhtar1971,Tanaka1972,Akhtar1975a,Naka1988}
It is therefore probable that the high Peierls stress of screw dislocations 
has an extrinsic cause. 
In pure zirconium or pure titanium,
this Peierls stress may not be as high and screw dislocations are probably gliding
as easily as other orientations.

Recently, Mendelev and Ackland\cite{Mendelev2007} developed
an empirical potential for Zr in the Embedded Atom Method (EAM) 
formalism.
Using this potential, Khater and Bacon\cite{Khater2010}
showed that it leads to a screw dislocation that spontaneously 
spreads in the prism plane, the configuration dissociated
in the basal plane being metastable.
They also showed that the Peierls stress of a screw dislocation 
gliding in the prism plane is not so different from the Peierls
stress of an edge dislocation and that this stress is small 
(22\,MPa for the screw and 16\,MPa for the edge).
These results therefore support experimental findings
stating that screw dislocations are gliding in prism planes
with a low Peierls stress in pure zirconium.
But these simulations rely on a central forces potential,
which is not well suited to describe dislocations in a hcp 
transition metal like Zr, as described above. 
More reliable atomistic simulations, 
incorporating a description of the electronic structure,
are therefore needed to confirm this easy glide of screw dislocations
in pure zirconium.

This article aims to use \abinitio calculations 
so as to fully characterize the core structure of a $1/3\langle1\bar{2}10\rangle$
screw dislocation in zirconium and estimate its Peierls stress.
We also examine generalized stacking faults as dislocation 
core structures are closely related to them.
Two different \abinitio methods, 
\siesta\cite{Soler2002} and \pwscf \cite{Giannozzi2009},
have been used.
\siesta offers the advantage of efficiency, allowing simulating
more atoms than with a standard \abinitio code, 
whereas \pwscf offers the advantage of robustness. 
All calculations are also performed with Mendelev and Ackland
EAM potential,\cite{Mendelev2007}
so as to identify its ability to model dislocations 
in zirconium. 
In addition, this empirical potential is used to study
the convergence of our results with the size of the simulation cell.

\section{Atomic interaction modeling}

Atomistic simulations have been performed both with an empirical interatomic potential
and with \abinitio calculations.
The empirical potential that we used is the EAM potential developed by Mendelev and Ackland.
\cite{Mendelev2007} 
This potential is labeled \#3 in Ref.~\onlinecite{Mendelev2007}. 
It is supposed to be well suited to model dislocations, 
as \abinitio values\cite{Domain2004a} of the stacking fault energies
in the basal and prism planes have been included in the fitting procedure.
Using this potential, Khater and Bacon \cite{Khater2010} showed
that a $1/3 \left< 1\bar{2}10\right>$ screw dislocation spontaneously 
dissociates in the prism plane and that a metastable configuration 
dissociated in the basal plane also exists.

The \abinitio calculations are relying on the Density Functional Theory
(DFT) in the Generalized Gradient Approximation (GGA) 
with the functional proposed by Perdew, Burke, and Ernzerhof (PBE)
and the pseudopotential approximation. 
Two different \abinitio codes are used, \siesta \cite{Soler2002} 
and \pwscf.\cite{Giannozzi2009} 

In the \siesta code,\cite{Soler2002} valence electrons are described by a localized basis set
corresponding to a linear combination of pseudoatomic orbitals with 13 functions per atom. 
We used a norm conserving pseudopotential of Troulliers-Martins type
with 4p electrons included as semicore.
Electronic density of state is broaden with the Methfessel-Paxton function
with a broadening of 0.3\,eV and the integration is performed on 
a regular grid of $14\times14\times8$ k-points for the primitive hcp unit cell
and an equivalent density of k-points for the supercells used for the defect calculations. 
The charge density is represented on a real space grid with a spacing of 0.08\,{\AA} 
(Mesh cutoff: 450\,Ry) that is reduced to 0.06\,{\AA} (800\,Ry) for dislocation calculations.
This approach, \ie the basis and the pseudopotential, has already been used to study 
vacancy diffusion in zirconium \cite{Verite2007a} and comparison with plane waves DFT 
calculations has led to a reasonable agreement.

In the \pwscf code,\cite{Giannozzi2009} valence electrons are described 
with plane waves using a cutoff energy of 28\,Ry.
The pseudopotential is ultrasoft of Vanderbilt type with 4s and 4p electrons included as semicore.
\footnote{Pseudopotential file Zr.pbe-nsp-van.UPF from www.quantum-espresso.org.}
The same k-point grid and the same electronic broadening are used as with \siesta code.

\begin{table}[hbt]
  \caption{Bulk properties of hcp Zr calculated with different atomic interaction models
  and compared to experimental data: 
  lattice parameter $a$, $c/a$ ratio of the hexagonal lattice,
  relaxed elastic constants $C_{ij}$,
  inner elastic constants \cite{Cousins1979} $e_{ij}$ and $d_{ij}$,
  phonon frequencies $\omega_1$ and $\omega_3$ of the optical branches at the $\Gamma$ point (Eq.~\ref{eq:phonon}),
  inner elasticity contribution to elastic constant $\delta C_{12}$ (Eq.~\ref{eq:dC12}).}
  \label{tab:bulk}
  \begin{ruledtabular}
  \begin{tabular}{lcccc}
    				& Expt. & EAM & \siesta & \pwscf \\
    \hline
    $a$ (\AA)			& 3.232\cite{Pearson1985}	& 3.234	& 3.237	& 3.230 \\
    $c/a$			& 1.603\cite{Pearson1985}	& 1.598	& 1.613	& 1.601 \\
    $C_{11}$ (GPa)		& 155.4\footnotemark[1]		& 142.	& 140.	& 140.	\\
    $C_{33}$ (GPa)		& 172.5\footnotemark[1]		& 168.	& 168.	& 168.	\\
    $C_{12}$ (GPa)		& 67.2\footnotemark[1]		& 75.	& 86.	& 70.	\\
    $C_{13}$ (GPa)		& 64.6\footnotemark[1]		& 76.	& 68.	& 65.	\\
    $C_{44}$ (GPa)		& 36.3\footnotemark[1]		& 44.	& 24.	& 26.	\\
    $C_{66}$ (GPa)		& 44.1\footnotemark[1]		& 33.5	& 27.	& 35.	\\
    $e_{11}$ (meV\ \AA$^{-5}$)	& 				& 27.0	& 17.0	& 18.6	\\
    $e_{33}$ (meV\ \AA$^{-5}$)	& 				& 118.	& 122.	& 101.	\\
    $d_{21}$ (meV\ \AA$^{-4}$)	& 				& 30.0	& 38.9	& 36.6	\\
    $\omega_1$ (THz)	& $2.66\pm0.02$\cite{Stassis1978}	& 2.60	& 2.08	& 2.16	\\
    $\omega_3$ (THz)	& $4.23\pm0.15$\cite{Stassis1978}	& 5.43 	& 5.57	& 5.03	\\
    $\delta C_{12}$ (GPa)	& 				& 5.33	& 14.3	& 11.5	
  \end{tabular}
  \end{ruledtabular}
  \footnotetext[1]{Experimental elastic constants\cite{Fisher1964} have been measured at 4\,K.}
\end{table}

To validate these different atomic interaction models, it is worth comparing their results
to available experimental data for some bulk properties of Zr.
All models lead to an equilibrium lattice parameter 
and a $c/a$ ratio in good agreement with experimental data (Tab.~\ref{tab:bulk}). 
In particular, a ratio lower than the ideal $\sqrt{8/3} \sim 1.633$ value is obtained in all cases.

We also compared the theoretical elastic constants with experimental data 
(Tab.~\ref{tab:bulk}): a good agreement is also obtained.
The computed elastic constants are the relaxed ones:\footnote{Elastic constants 
calculated by Mendelev and Ackland in the original article 
describing the EAM potential\cite{Mendelev2007} did not take into account 
atomic relaxations. This explains the difference with the ones given in Tab.~\ref{tab:bulk}.}
as the hcp lattice contains two atoms in its primitive unit cell, 
some internal degrees of freedom may exist when applying a homogeneous strain.
One needs to allow for atomic relaxations when computing $C_{11}$, $C_{12}$ or $C_{66}$
constants.\cite{Cousins1979}
It is also possible to calculate inner elastic constants to characterize these internal degrees of freedom.
These are also given in Tab.~\ref{tab:bulk} using the notations introduced by Cousins.\cite{Cousins1979}
Two of these inner elastic constants, $e_{11}$ and $e_{33}$, are related to the 
phonon frequencies of the optical branches at the $\Gamma$ point\cite{Cousins1979}
\begin{equation}
  \omega_i = 2 \sqrt{ \frac{\Omega e_{ii}}{m}},
  \label{eq:phonon}
\end{equation}
where $\Omega = a^2 c \sqrt{3}/4$ is the atomic volume and $m$ the atomic mass.
The last inner elastic constants $d_{21}$ couples the internal degrees of freedom
to the homogeneous strain. It leads to a contribution to the elastic constants\cite{Cousins1979}
\begin{equation}
  \delta C_{12} = \frac{ {d_{21}}^2}{e_{11}}.
  \label{eq:dC12}
\end{equation}
If $C^0_{ij}$ are the unrelaxed elastic constants, \ie the elastic constants
calculated by imposing a homogeneous strain to the hcp lattice without letting 
atoms relax from their initial positions, the true elastic constants are given by\cite{Cousins1979}
$C_{11}=C^0_{11}-\delta C_{12}$, $C_{12}=C^0_{12}+\delta C_{12}$, and $C_{66}=C^0_{66}-\delta C_{12}$,
all other elastic constants being unchanged.

\section{Stacking fault energies}

Dislocation dissociation is controlled by the existence
of a metastable stacking fault for the corresponding plane. 
According to the results obtained by Khater and Bacon \cite{Khater2010} 
with Mendelev and Ackland EAM potential,\cite{Mendelev2007}
a $1/3\left<1\bar{2}10\right>$ screw dislocation can dissociate either 
in a basal or in a prism plane.
To characterize these eventual dissociations, 
we compute generalized stacking fault energies \cite{Vitek1968,Vitek2008}
-- or $\gamma$-surfaces --
for both the basal and the prism planes. 

\subsection{Methodology}

$\gamma$-surfaces describe the energy variation when two parts of a crystal 
are rigidly shifted for different fault vectors lying in a given crystallographic plane.
Atoms are allowed to relax in the direction perpendicular to the fault plane.
We calculate these $\gamma$-surfaces for both the basal and prism planes 
using full periodic boundary conditions. To introduce only one fault
in the simulation cell, the same shift as the one corresponding to the fault vector 
is applied to the periodicity vector perpendicular to the fault plane.
No free surfaces is therefore introduced in the simulation cell, which allows
a fast convergence of the fault energies with the number of stacked planes.
A periodic stacking of at least 16 $\left\{ 0001 \right\}$ planes is used for the basal fault
and 12 $\left\{ 10\bar{1}0 \right\}$ planes for the prismatic fault.
This corresponds to a distance between fault planes $h_{0001}=8c\sim41$\,{\AA}
and $h_{10\bar{1}0}=6\sqrt{3}a\sim34$\,{\AA}.
Increasing the number of planes in the stacking modifies the energies
by less than 1\,mJ\,m$^{-2}$.
Generalized stacking fault energies are calculated on a regular grid
of $10\times10$ fault vectors and are then interpolated with Fourier series.

\subsection{$\gamma$-surfaces}

\begin{table}
  \caption{Stacking fault energies in the basal plane, $\gamma_{\rm b}$,
  and in the prism plane, $\gamma_{\rm p}$, calculated with different 
  atomic interaction models, including \vasp calculations 
  of Domain \etal.\cite{Domain2004a}
  and of Udagawa \etal.\cite{Udagawa2010}
  $R$ is the ratio defined by Legrand,\cite{Legrand1984} 
  and $d^{\rm eq}_{\rm b}$ and $d^{\rm eq}_{\rm p}$ are the dissociation lengths
  of a screw dislocation respectively in the basal (Eq.~\ref{eq:dBasal})
  and in the prism plane (Eq.~\ref{eq:dPrismatic}).}
  \label{tab:gFault}
  \begin{ruledtabular}
  \begin{tabular}{lccccc}
    			& EAM	& \siesta	& \pwscf	& \vasp\cite{Domain2004a}	& \vasp\cite{Udagawa2010}	\\
    \hline
    $\gamma_{\rm b}$ 
    (mJ\,m$^{-2}$) 	& 198.	& 199.		& 213.		& 200.		& 227.	\\
    $\gamma_{\rm p}$ (mJ\,m$^{-2}$) 	
    			& 135. (274.)\footnotemark[1]	
    				& 233.		& 211.		& 145.		& 197.	\\
    $R = C_{66}\gamma_{\rm b} / C_{44}\gamma_{\rm p}$	
    			& 1.12	& 0.96		& 1.36		& 1.85		& 2.1	\\
    $d^{\rm eq}_{\rm b}$ (\AA)
    			& 4.0	& 2.0		& 2.7		& 3.4		& 3.2	\\
    $d^{\rm eq}_{\rm p}$ (\AA)			
    			& 7.8	& 4.6		& 5.9		& 9.6 		& 7.4	\\
  \end{tabular}
  \end{ruledtabular}
  \footnotetext[1]{For the EAM calculation of the prismatic stacking fault energy,
  the value in parenthesis corresponds to the maximum in $a/6[1\bar{2}10]$, 
  whereas the lower value corresponds to the true minimum in $a/6\ [1\bar{2}10]+0.14c \ [0001]$.}
\end{table}

\subsubsection{Basal plane}

\begin{figure}[hbtp]
    \includegraphics[width=0.9\linewidth]{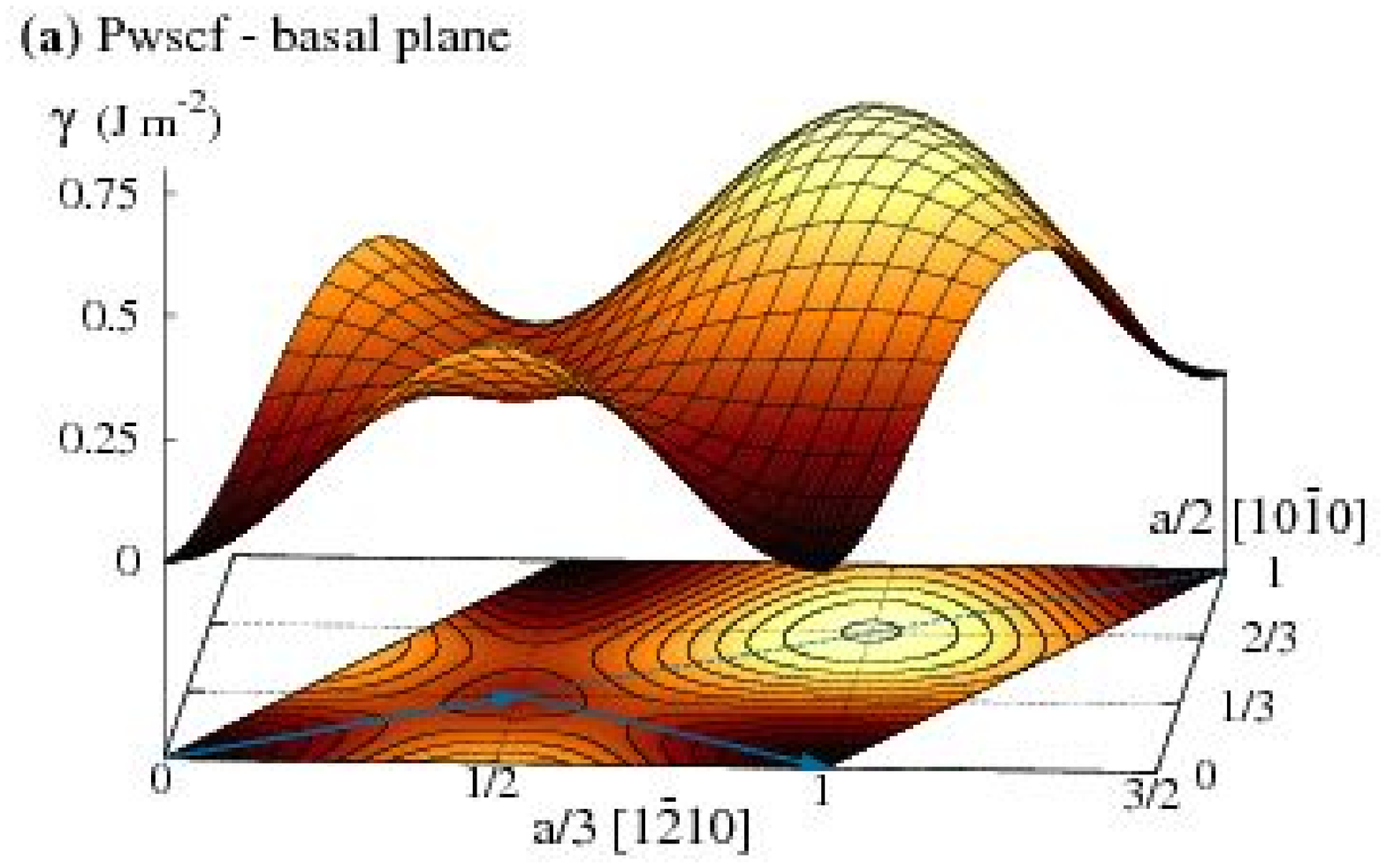}
    \includegraphics[width=0.9\linewidth]{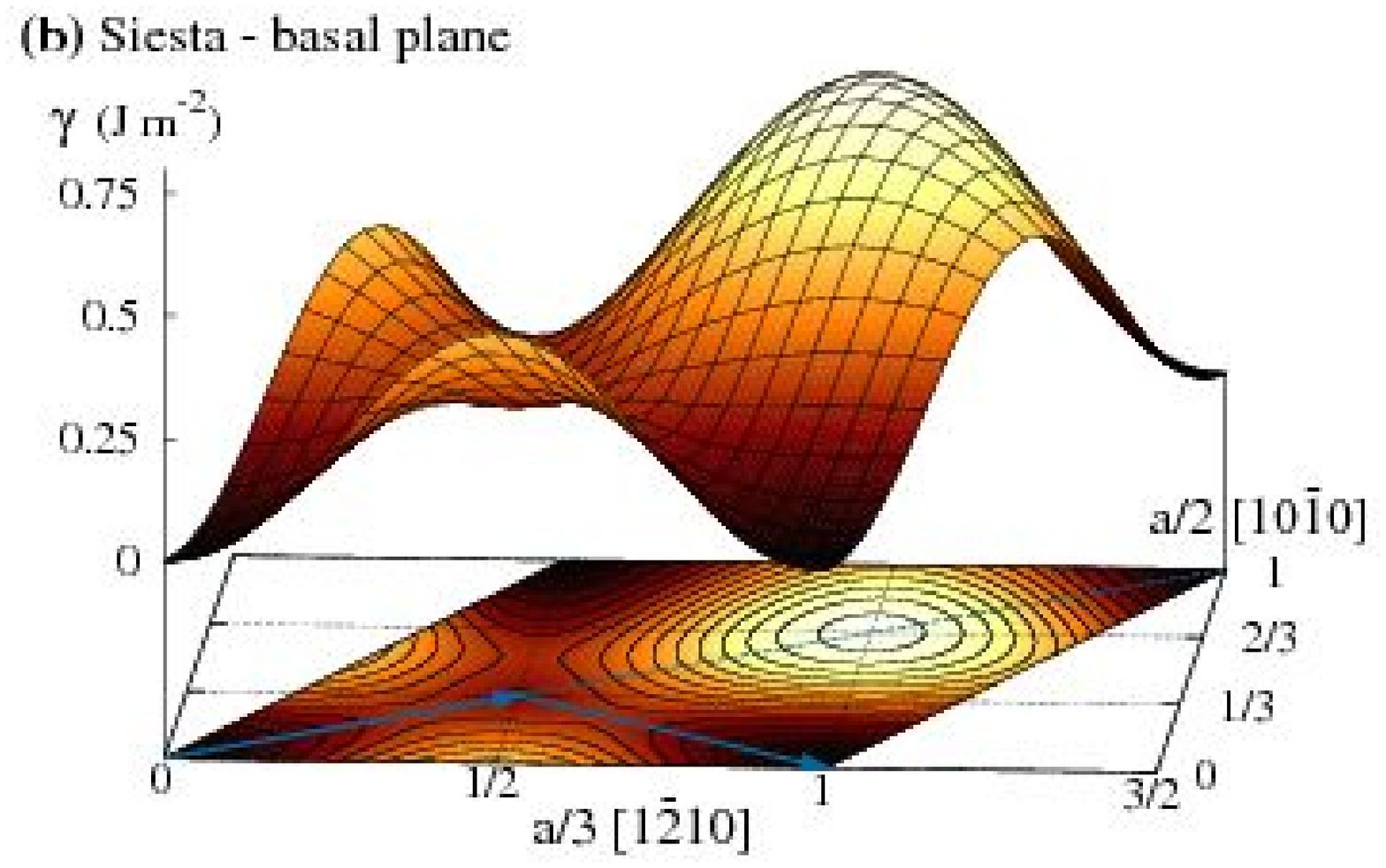}
    \includegraphics[width=0.9\linewidth]{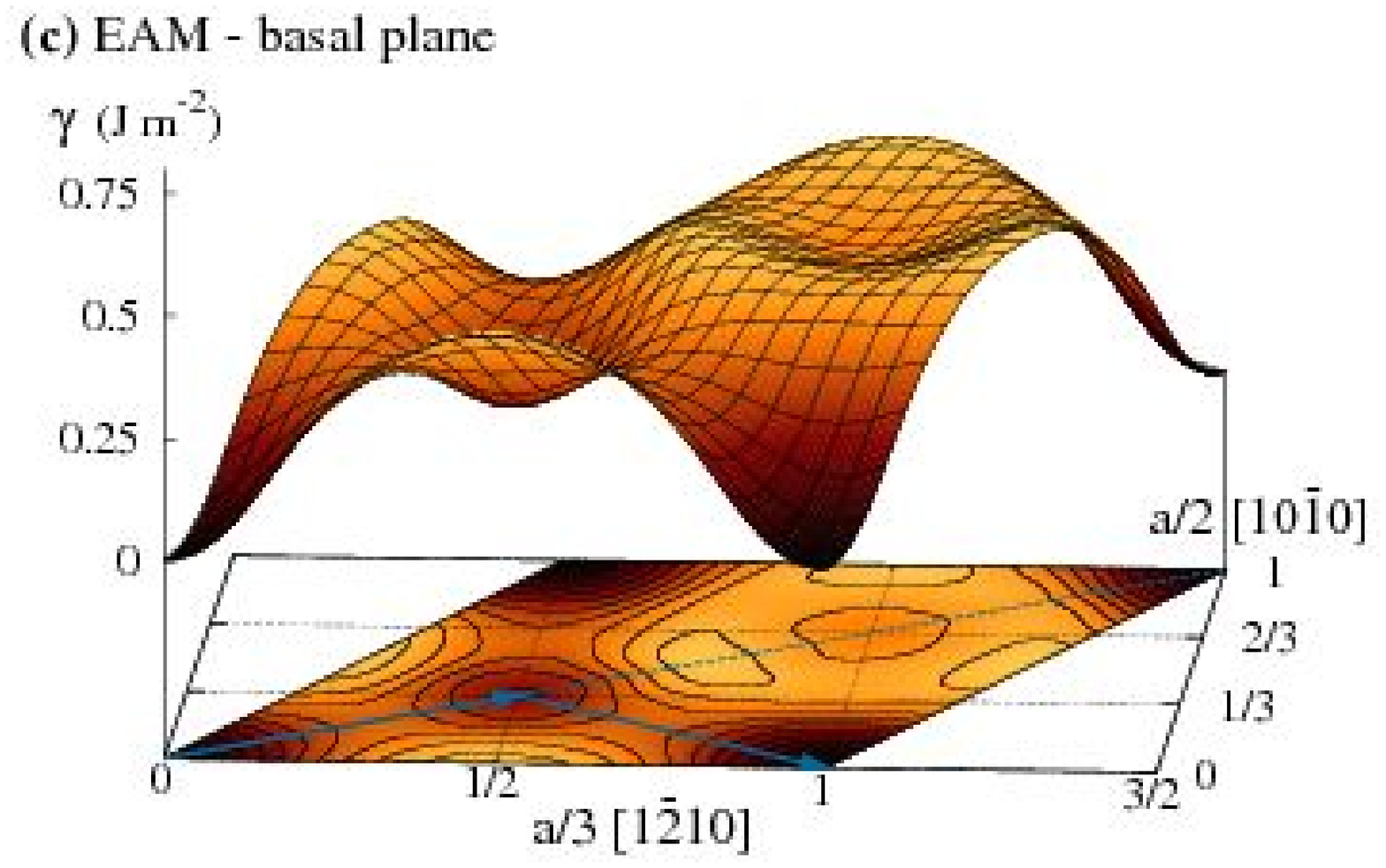}
  \caption{Generalized stacking fault energy in the basal plane 
  	calculated with (a) \pwscf, (b) \siesta and (c) EAM potential.
	The arrows indicate Burgers vectors of the partial dislocations
	corresponding to a dissociation in the basal plane.
	The dashed line is the $[1\bar{1}00]$ direction used in Fig.~\ref{fig:gLineBasal}.
	Contour lines are drawn at the base every 50\,mJ\,m$^{-2}$.}
  \label{fig:gSurfaceBasal}
\end{figure}

\begin{figure}[hbtp]
    \includegraphics[width=0.9\linewidth]{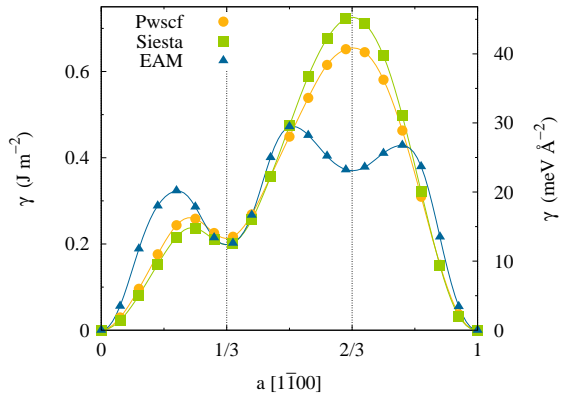}
  \caption{Generalized staking fault energy in the basal plane 
  along the $[1\bar{1}00]$ direction (\cf Fig.~\ref{fig:gSurfaceBasal})
  calculated with \pwscf, \siesta and EAM potential.}
  \label{fig:gLineBasal}
\end{figure}

$\gamma$-surfaces corresponding to the basal plane are shown in Fig.~\ref{fig:gSurfaceBasal}
for the three interaction models we used.
In all cases, a local minimum can be found at $1/3[1\bar{1}00]$
which corresponds to the intrinsic $I_2$ fault.\cite{Hull2001}
This minimum does not vary when full atomic relaxations are allowed
instead of being constrained to the direction perpendicular to the fault plane.
All methods give a close value for the fault energy $\gamma_{\rm b}$ in this minimum
(Tab.~\ref{tab:gFault}). A good agreement is also obtained with the values 
calculated by Domain \etal \cite{Domain2004a}
and by Udagawa \etal \cite{Udagawa2010} using \vasp \abinitio code.
The depth of this local minimum is more pronounced with the empirical
EAM potential [Fig.~\ref{fig:gSurfaceBasal}(c)]
than in \abinitio calculations [Fig.~\ref{fig:gSurfaceBasal}(a) and (b)].
This appears clearly in Fig.~\ref{fig:gLineBasal} where the fault 
energy predicted by the different interaction models are compared 
along the $[1\bar{1}00]$ direction.
We will see latter that this has consequences on the stability 
of a screw dislocation dissociated in the basal plane.

The $\gamma$-surface calculated with the EAM empirical potential 
has another minimum located in $2/3[1\bar{1}00]$.
This is an artifact of the potential:
one expects instead a maximum as this fault vector transforms the original 
BABABA stacking of the basal planes in a BABBCB stacking.
\Abinitio calculations confirm that this fault vector gives a maximum.
Finally, it is worth pointing that both \abinitio techniques give 
a very similar $\gamma$-surface: the shapes are identical and the amplitudes
do not differ by more than 10\%.

\subsubsection{Prism plane}

\begin{figure}[hbtp]
    \includegraphics[width=0.9\linewidth]{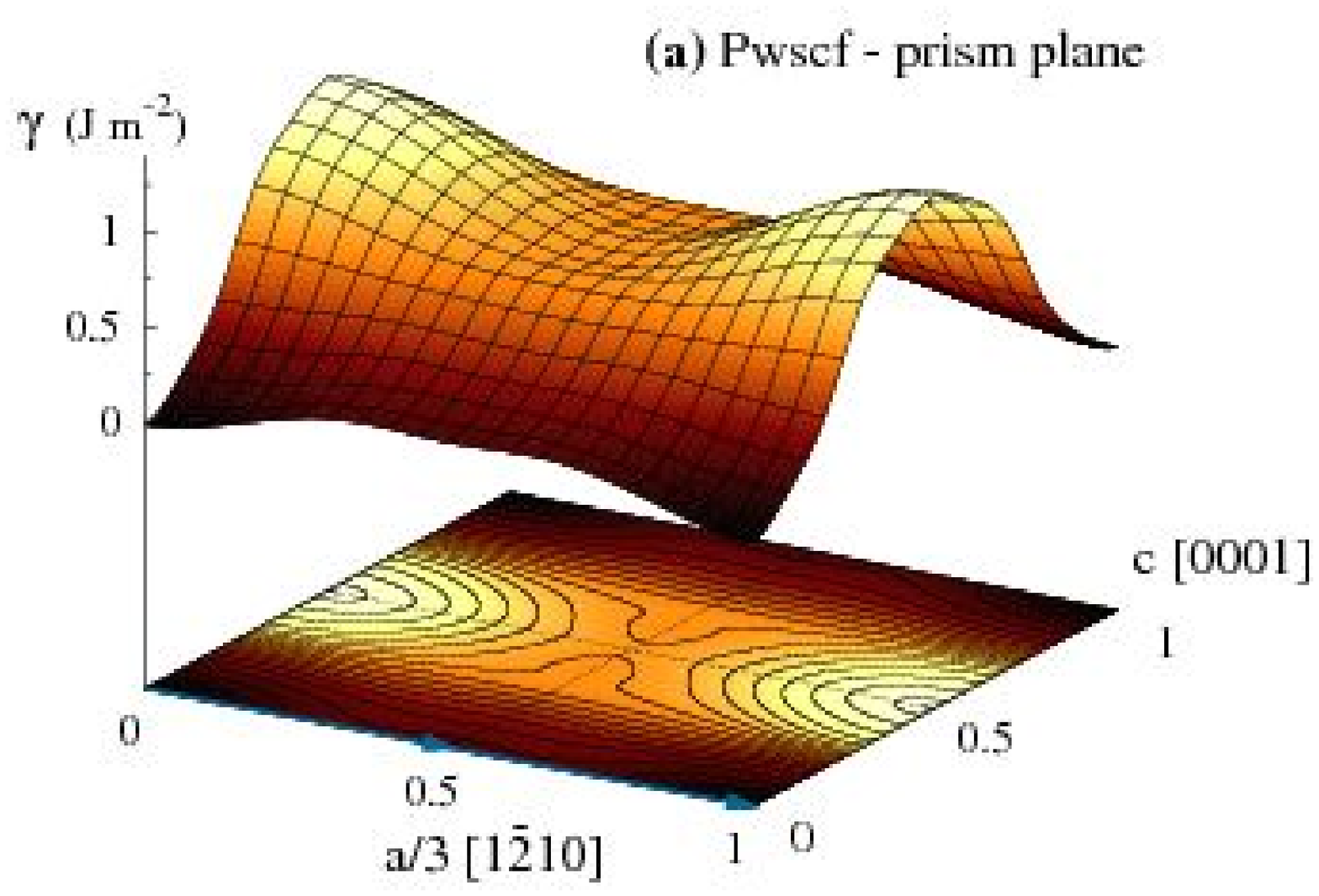}
    \includegraphics[width=0.9\linewidth]{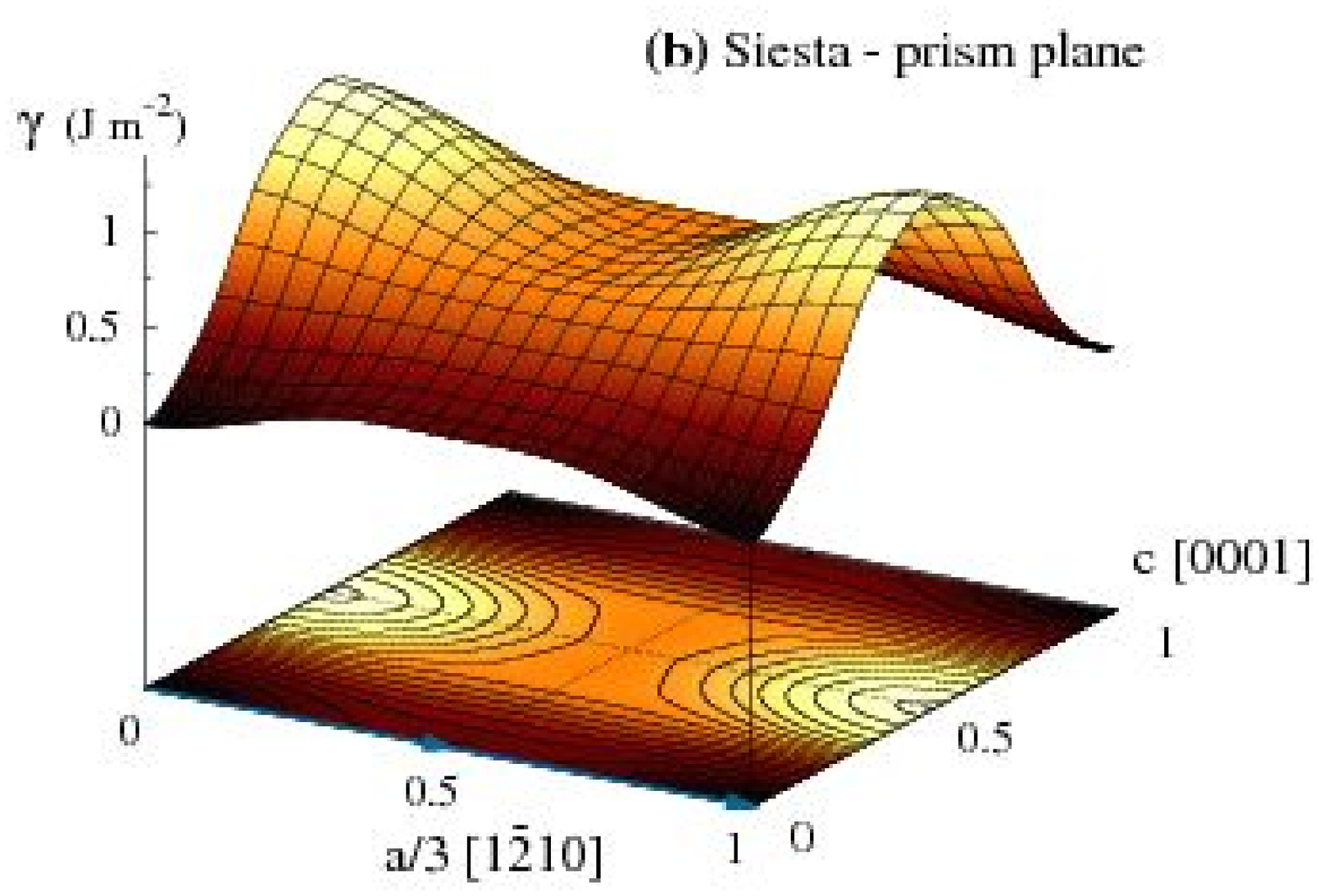}
    \includegraphics[width=0.9\linewidth]{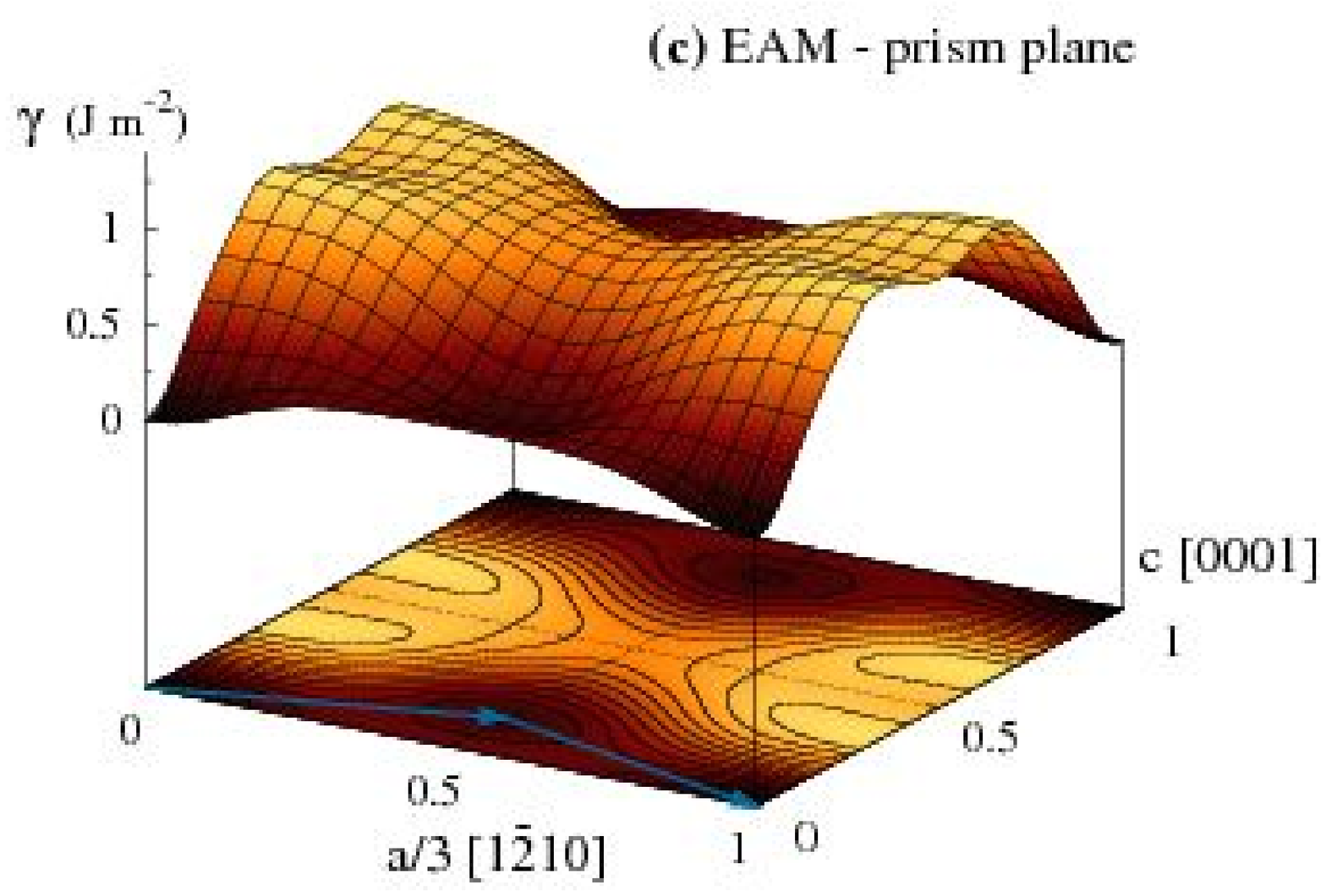}
  \caption{Generalized stacking fault energy in the prism plane 
  	calculated with (a) \pwscf, (b) \siesta and (c) EAM potential.
	The arrows indicate Burgers vectors of the partial dislocations
	corresponding to a dissociation in the prism plane.
	Contour lines are drawn at the base every 75\,mJ\,m$^{-2}$.}
  \label{fig:gSurfacePrismatic}
\end{figure}

\begin{figure}[hbtp]
    \includegraphics[width=0.9\linewidth]{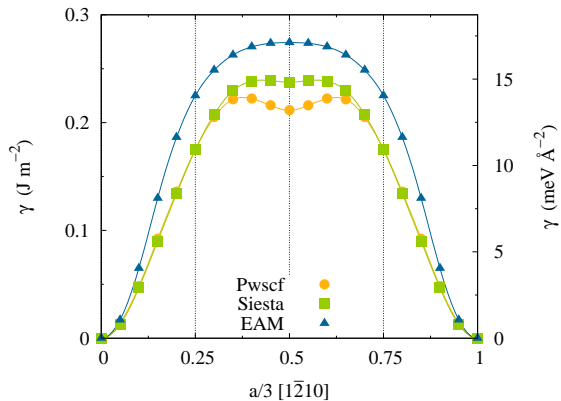}
  \caption{Generalized staking fault energy in the prism plane 
  along the $[1\bar{2}10]$ direction
  calculated with \pwscf, \siesta and EAM potential.}
  \label{fig:gLinePrismatic}
\end{figure}

$\gamma$-surfaces for the prism plane are shown in Fig.~\ref{fig:gSurfacePrismatic}.
Both \abinitio methods lead to a similar $\gamma$-surface [Figs.~\ref{fig:gSurfacePrismatic}(a) and (b)].
In particular, both \pwscf and \siesta predicts the existence of a minimum 
at halfway of the Burgers vector, \ie in $1/6\ [1\bar{2}10]$. 
Like for the basal fault, this minimum does not vary when full atomic relaxations are allowed.
As can be seen on the projection of these $\gamma$-surface along the $[1\bar{2}10]$ direction
(Fig.~\ref{fig:gLinePrismatic}), this minimum is a little more pronounced with \pwscf than with \siesta.
The same minimum was also present in the \vasp calculations of Domain \etal,\cite{Domain2004a}
but they obtained a lower value $\gamma_{\rm p}$ of the fault energy in this point 
(Tab.~\ref{tab:gFault}).
On the other hand, Udagawa \etal \cite{Udagawa2010} obtained a value 
close to our result using also \vasp \abinitio code with the PAW method
and the PBE-GGA functional. They pointed out that the discrepancy 
arises from an insufficient number of stacked planes in the $\gamma$-surface
calculation of Domain \etal.
The energy of the metastable stacking fault energy in the prism plane 
appears therefore higher than the value 145\,mJ\,m$^{-2}$ initially suggested by Domain \etal:\cite{Domain2004a}
both our \pwscf\footnote{See Supplemental Material at
http://link.aps.org/supplemental/10.1103/PhysRevB.86.144104
for a discussion on the convergence of this fault energy with \pwscf parameters.} and \siesta calculations, 
as well as the Udagawa \etal result,\cite{Udagawa2010} leads to a value of about 200\,mJ\,m$^{-2}$.

The $\gamma$-surface calculated with the EAM potential is quite different 
[Fig.~\ref{fig:gSurfacePrismatic}(c)]
as the point located in $1/6\ [1\bar{2}10]$ is indeed a maximum 
and not a minimum like with \abinitio calculations.
A minimum is found for a point located in $a/6\ [1\bar{2}10]+0.14c \ [0001]$.
One therefore expects that a $1/3 \langle1\bar{2}10\rangle \left\{10\bar{1}0\right\}$ dislocation
dissociates into two partial dislocations with a Burgers vector component orthogonal 
to the one of the perfect dislocation. In particular, a screw dislocation should dissociate
in two partial dislocations with a small edge character.
Khater and Bacon \cite{Khater2010} showed that the empirical potential 
of Ackland \etal \cite{Ackland1995} suffers from the same artefact.
As noted by Bacon and Vitek,\cite{Bacon2002} all central forces potentials
stabilize indeed such a stacking fault
$a/6\ [1\bar{2}10] + \alpha c \ [0001]$ with $\alpha \neq 0$,
a minimum also predicted by a simple hard sphere model.\cite{Schwartzkopff1969}
This minimum either disappears or is located exactly in $a/6\ [1\bar{2}10]$
($\alpha=0$) only when the angular dependence of the atomic interactions
is taken into account.
The value $\gamma_{\rm p}$ of the stacking fault energy 
obtained with Mendelev and Ackland EAM potential \cite{Mendelev2007} for this minimum 
is much lower than our \abinitio value (Tab.~\ref{tab:gFault}). 
This is quite normal as Mendelev and Ackland used the \abinitio value
of Domain \etal to fit their potential.


\subsection{Dislocation dissociation}
\label{sec:dissociation}

Before using atomistic simulations to obtain dislocation core structures 
and their associated Peierls stress, it is worth looking what can be learned 
from these $\gamma$-surface calculations.
Legrand \cite{Legrand1984} proposed a criterion based on the elastic constants
and the metastable stacking fault energies to determine if glide occurs in the 
base or prism plane in a hcp crystal. 
According to this criterion, prismatic glide is favored if the ratio 
$R = C_{66} \gamma_{\rm b} / C_{44} \gamma_{\rm p}$ is larger than 1.
Tab~\ref{tab:gFault} shows that this is the case for the EAM potential
and the \pwscf calculation, as well as for the \vasp calculation
of Domain \etal\cite{Domain2004a} and of Udagawa \etal.\cite{Udagawa2010}
On the other hand, \siesta leads to a value 
too close to 1 to be able to decide between basal and prismatic glide.

One can also use dislocation elasticity theory \cite{Hirth1982} 
to compute the dissociation distance
of a dislocation both in the basal and prism planes.
According to elasticity theory, the energy variation caused 
by a dissociation of length $d$ is 
\begin{equation}
  \Delta E_{\rm diss}(d) = -b_i^{(1)}K_{ij}b_j^{(2)} \ln{\left(\frac{d}{r_{\rm c}}\right)} + \gamma d,
  \label{eq:Edissociation}
\end{equation}
where $\mathbf{b}^{(1)}$ and $\mathbf{b}^{(2)}$ are the Burgers vectors of each partial dislocation, 
$\gamma$ the corresponding stacking fault energy,
$K$ the Stroh matrix\cite{Stroh1958,Stroh1962} controlling dislocation elastic energy,
and $r_{\rm c}$ the core radius.
Minimization of this energy leads to the equilibrium dissociation length
\begin{equation}
  d^{\rm eq} = \frac{b_i^{(1)}K_{ij}b_j^{(2)}}{\gamma}.
  \label{eq:dissociation}
\end{equation}
When the hcp crystal is oriented with the $x$, $y$, and $z$ axis respectively along the
$[10\bar{1}0]$, $[0001]$, and $[1\bar{2}10]$ directions, 
for a dislocation lying along the $z$ direction,
the $K$ matrix is diagonal with its components given by\cite{Foreman1955,Teutonico1970,Savin1976}
\footnote{Foreman \cite{Foreman1955} gave a different expression for $K_{22}$, but the comparison
with a numerical evaluation of the $K$ matrix using Stroh formalism \cite{Stroh1958,Stroh1962} 
shows that the correct expression is the one given by Savin \etal.\cite{Savin1976}}
\begin{equation*}
  \begin{split}
    K_{11} &= \frac{1}{2\pi} \left( \bar{C}_{11} + C_{13} \right) 
    \sqrt{\frac{C_{44}\left( \bar{C}_{11}-C_{13} \right)}{C_{33}\left( \bar{C}_{11}+C_{13}+2C_{44} \right)}}, \\
    K_{22} &= \sqrt{ \frac{C_{33}}{C_{11}} } K_{11}, \\
    K_{33} &= \frac{1}{2\pi} \sqrt{\frac{1}{2} C_{44}\left( C_{11} - C_{12} \right)},
  \end{split}
\end{equation*}
where $\bar{C}_{11}=\sqrt{C_{11}C_{33}}$.

The $\gamma$-surface of the basal plane (Fig.~\ref{fig:gSurfaceBasal})
indicates a possible dissociation 
$1/3[1\bar{2}10] \to 1/3[1\bar{1}00] + 1/3[0\bar{1}10]$.
The dissociation length in the basal plane is then, for a $1/3[1\bar{2}10]$ screw dislocation,
\begin{equation}
  d^{\rm eq}_{\rm b} = \frac{\left( 3 K_{33}-K_{11} \right)a^2}{12 \gamma_{\rm b}}.
  \label{eq:dBasal}
\end{equation}

According to the minimum of the $\gamma$-surface in the prism plane (Fig.~\ref{fig:gSurfacePrismatic}), 
a $1/3[1\bar{2}10]$ dislocation can dissociate in this plane in two partial dislocations 
with Burgers vectors $1/6[1\bar{2}10] \pm \alpha c/a [0001]$.
The parameter $\alpha$ controls the position of the stacking fault minimum along the $[0001]$ direction,
\ie $\alpha=0$ for \pwscf and \siesta, and $\alpha=0.14$ for the EAM potential.
The dissociation length of a screw dislocation in the prism plane is then
\begin{equation}
  d^{\rm eq}_{\rm p} = \frac{\left( K_{33} a^2 - 4 \alpha^2 K_{22} c^2 \right)}{4 \gamma_{\rm p}}.
  \label{eq:dPrismatic}
\end{equation}

The dissociation lengths $d^{\rm eq}_{\rm b}$ and $d^{\rm eq}_{\rm p}$ calculated from the elastic constants
and the stacking fault energies are given in table \ref{tab:gFault}.
Elastic constants predicted by the atomic interaction models are used in each case.
For all energy models, one expects a larger dissociation in the prism than
in the basal plane.
We will compare in the following section these dissociation lengths predicted by elasticity 
theory with the ones observed in our atomistic simulations of the dislocation core structure.

\section{Screw dislocation core}
\label{sec:1:dislo}

\subsection{Methodology}

\begin{figure}[tbh]
    \includegraphics[width=0.99\linewidth]{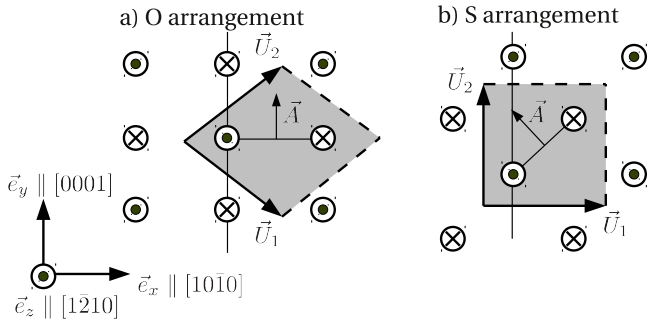}
  \caption{Screw dislocation periodic arrangements used for atomistic simulations. 
  $\vec{U}_1$ and $\vec{U}_2$ are the periodicity vectors of the arrangement,
  and $\vec{A}$ the cut vector of the dislocation dipole.
  The thin vertical line corresponds to the prism glide plane.}
  \label{fig:periodicArray}
\end{figure}

Our atomistic simulations of the core structure of a screw dislocation 
are based on full periodic boundary conditions.\cite{Bulatov2006}
This requires introducing a dislocation dipole in the simulation cell.
Two periodic arrangements of the dislocations have been used (Fig.~\ref{fig:periodicArray}).
In the O arrangement, dislocations with opposite Burgers vectors are located 
on the same prism and basal planes, \ie the two foreseen glide planes.
On the other hand, only dislocation with the same Burgers vectors can be found 
on a given prism or basal plane in the S arrangement.

Periodicity vectors of the O arrangement are, before introducing the dislocations,
$\vec{U}_1 = n a \frac{1}{2}\left[ 10\bar10 \right] - m c \left[ 0001 \right]$, 
$\vec{U}_2 = n a \frac{1}{2}\left[ 10\bar10 \right] + m c \left[ 0001 \right]$, and
$\vec{U}_3 = a \frac{1}{3} \left[ 1\bar{2}10 \right]$.
The integers $n$ and $m$ are taken equal to keep an aspect ratio close to a square.
This simulation cell has been used for \abinitio calculations with 
$n=4$ (128 atoms),
$n=5$ (160 atoms), and
$n=6$ (288 atoms).

For the S arrangement, 
$\vec{U}_1 = n a \frac{1}{2}\left[ 10\bar10 \right]$, 
$\vec{U}_2 = m c \left[ 0001 \right]$, and
$\vec{U}_3 = a \frac{1}{3} \left[ 1\bar{2}10 \right]$.
\Abinitio calculations have been performed with 
$n=m$ and varying between $n=5$ (100 atoms) and
$n=8$ (256 atoms).

Both dislocation arrays are quadrupolar: the vector joining the two dislocations 
composing the primitive dipole is $\vec{D} = 1/2( \vec{U}_{1}+\vec{U}_2 )$.
Because of the centrosymmetry of this arrangement 
and the symmetry properties of the Volterra elastic field, 
this ensures that the stress created by other dislocations is minimal 
at each dislocation position.
The cut vector $\vec{A}$ defining the dislocation dipole is obtained
by a $\pi/2$ rotation of $\vec{D}$.
 
The dislocation dipole is introduced in the simulation cells by applying to all atoms
the elastic displacement predicted by anisotropic elasticity theory \cite{Stroh1958,Stroh1962} 
taking full account of the periodic boundary conditions.\cite{Cai2003}
A homogeneous strain is also applied to the simulation cell so as to cancel 
the plastic strain introduced by the dislocation dipole and minimize
the elastic energy. This strain is given by
\begin{equation*}
  \varepsilon^0_{ij} = - \frac{b_i A_j + b_j A_i}{2S},
\end{equation*}
where $S = | (\vec{U}_1 \wedge \vec{U}_2 ) \cdot \vec{e}_z |$ 
is the surface of the simulation cell perpendicular to the dislocation lines.
This homogeneous strain adds some tilt components to the periodicity vectors.
Atoms are then relaxed until all components of the atomic forces 
are smaller than 5\ meV\ \AA$^{-1}$ for \siesta,
2\ meV\ \AA$^{-1}$ for \pwscf,
and 0.1\ meV\ \AA$^{-1}$ for the EAM potential.

\subsection{Core structure}
\label{sec:2:core}

\begin{figure}[bth]
    \includegraphics[height=0.50\linewidth]{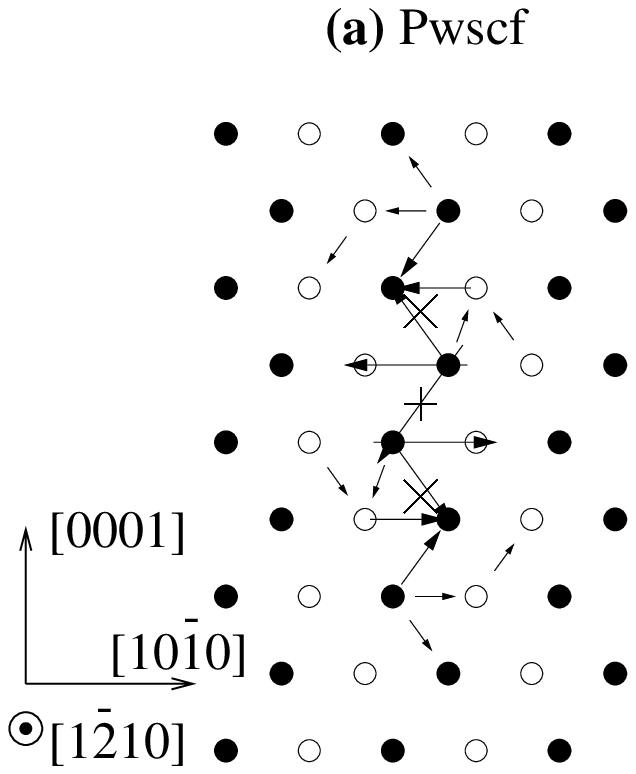}
    \hfill
    \includegraphics[height=0.50\linewidth]{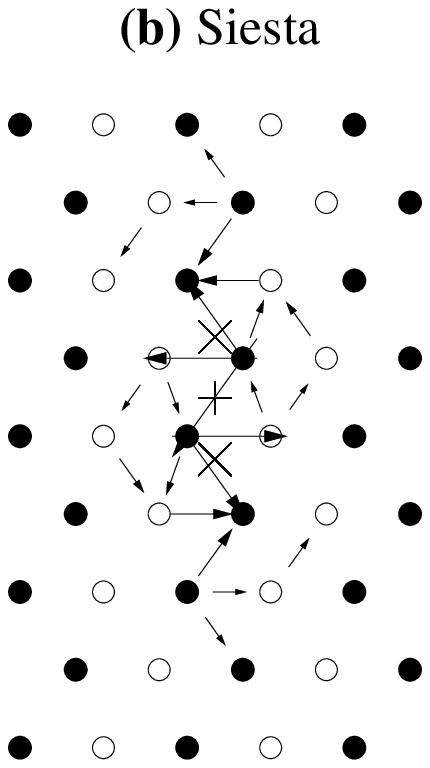}
    \hfill
    \includegraphics[height=0.50\linewidth]{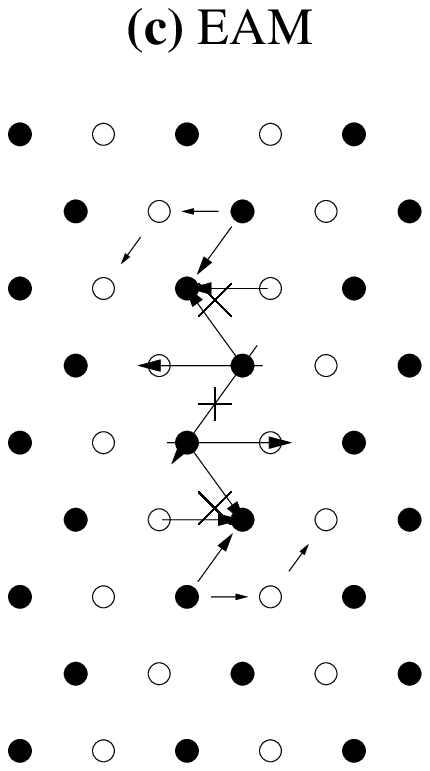}
  \caption{Differential displacement maps around one of the two $1/3[1\bar{2}10]$
  screw dislocations composing the dipole 
  for the S periodic arrangement with $n=m=7$ (196 atoms).
  Atoms are sketched by circles with a color depending on the $(1\bar{2}10)$ plane to which they belong.
  The arrow between two atomic columns is proportional to the $[1\bar{2}10]$ component of the differential displacement 
  between the two atoms. 
  Displacements smaller than $0.1 b$ are not shown.
  Crosses $\times$ correspond to the positions of the two partial dislocations,
  and $+$ to their middle, \ie the position of the total dislocation.}
  \label{fig:vitek}
\end{figure}

Starting from perfect dislocations, atom relaxation 
leads to dislocations spread in the prism plane,
whatever the interaction model (EAM, \siesta, or \pwscf)
and whatever the simulation cell used. 
This can be clearly seen by plotting differential displacement maps 
as introduced by Vitek.\cite{Vitek1970}
These maps (Fig.~\ref{fig:vitek}) show that the strain created by the screw dislocation 
spreads out in the $(10\bar{1}0)$ prism plane
and that displacements outside this plane are much smaller.

\begin{figure}[bth]
    \includegraphics[width=0.99\linewidth]{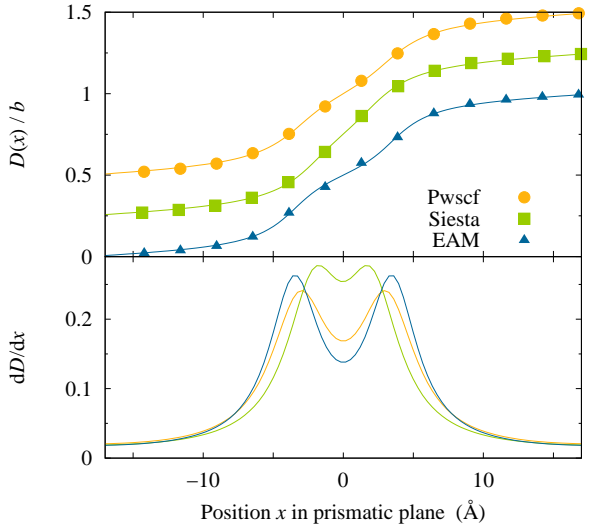}
  \caption{Disregistry $D(x)$ created by the screw dislocation in the prism plane
  and corresponding dislocation density $\rho(x)=\partial D(x) / \partial x$
  for the S periodic arrangement with 196 atoms ($n=m=7$).
  Symbols correspond to atomistic simulations and lines to the fit of the Peierls-Nabarro 
  model to these data.
  For clarity, disregistries $D(x)$ have been shifted by 0.25 between the different interaction models.}
  \label{fig:disregistry}
\end{figure}

To characterize this spreading in the $(10\bar{1}0)$
prism plane, we extract from our atomistic simulations 
the disregistry $D(x)$ created by the dislocation.
This is defined as the displacement difference between 
the atoms in the plane just above and those just below the dislocation 
glide plane. The derivative of this function, $\rho(x)=\partial D/\partial x$
corresponds to the dislocation density.
Fig.~\ref{fig:disregistry} shows the disregistry obtained 
for the three interaction models. In all cases, 
the $b$ discontinuity created by the screw dislocation 
does not show a sharp interface, but spreads on a distance
$\sim 10$\,\AA.

Peierls \cite{Peierls1940} and Nabarro \cite{Nabarro1947} built a model 
that leads to a simple expression of the disregistry. 
The analytical expression they obtained \cite{Hirth1982} can be extended 
to a dissociated dislocation. 
As suggested by the prismatic $\gamma$-surface (Fig.~\ref{fig:gSurfacePrismatic}), 
we assume that the screw dislocation dissociates in two equivalent partial dislocation
separated by a distance $d$.
Based on the Peierls-Nabarro model, we write the disregistry 
created by a single dislocation  
\begin{equation}
  \begin{split}
    D_{\rm dislo}(x) =& \frac{b}{2\pi}\left\{ \arctan{\left[\frac{x-x_0-d/2}{\zeta}\right]} \right.\\
  & \left. + \arctan{\left[\frac{x-x_0+d/2}{\zeta}\right]}
    + \pi/2 \right\},
  \end{split}
  \label{eq:PN}
\end{equation}
where $x_0$ is the dislocation position,
$d$ its dissociation length, 
and $\zeta$ the spreading of each partial dislocation.
We need then to take into account that we do not have only one dislocation 
on a given prism plane but a periodic array (Fig.~\ref{fig:periodicArray}).
The disregistry created by an array of period $L$ is
\begin{equation}
  \begin{split}
    D_L(x) =& \sum_{n=-\infty}^{\infty}{D_{\rm dislo}(x-nL)} \\
                  =& \frac{b}{2\pi}\left\{ 
		  \arctan{ \left[ \frac{\tan{\left(\frac{\pi}{L}[x-x_0-d/2]\right)}} {\tanh{\left(\frac{\pi\zeta}{L}\right)}} \right] } 
		  \right.\\
		  & \left. 
		  + \pi \bigg\lfloor \frac{x-x_0-d/2}{L} + \frac{1}{2} \bigg\rfloor
		  \right.\\
		  & \left. 
		  + \arctan{ \left[ \frac{\tan{\left(\frac{\pi}{L}[x-x_0+d/2]\right)}} {\tanh{\left(\frac{\pi\zeta}{L}\right)}} \right] } 
		  \right.\\
		  & \left. 
		  + \pi \bigg\lfloor \frac{x-x_0+d/2}{L} + \frac{1}{2} \bigg\rfloor
		  \right\},
  \end{split}
  \label{eq:PNperio}
\end{equation}
where $\lfloor\cdot\rfloor$ is the floor function.
For the O arrangement (Fig.~\ref{fig:periodicArray}a), the disregistry 
in the prism plane should be given by $D(x)=D_L(x)-D_L(x-L/2)$ with $L=2mc$,
whereas it should be $D(x)=D_L(x)$ with $L=mc$ for the S arrangement 
(Fig.~\ref{fig:periodicArray}b).

We fit this analytical expression of the dislocation disregistry 
to the data coming from the atomistic simulations. 
Fig.~\ref{fig:disregistry} shows a good agreement between atomistic simulations
and the model, using only three fitting parameters:
the dislocation position $x_0$,
the dissociation length $d$ and the spreading $\zeta$.
This procedure therefore allows us to determine the location of the dislocation center.
For all interaction models, we find that this center lies
in between two $(0001)$ atomic planes. 
One can see in Fig.~\ref{fig:vitek} that this position 
corresponds to a local symmetry axis of the differential displacement map.
This is different from the result obtained by Ghazisaeidi and Trinkle\cite{Ghazisaeidi2012} in Ti
where the center of the screw dislocation was found to lie exactly in 
one $(0001)$ atomic plane, a position that corresponds in Zr to the saddle point
between two Peierls valleys, as it will be shown below.

\begin{figure}[bth]
    \includegraphics[width=0.9\linewidth]{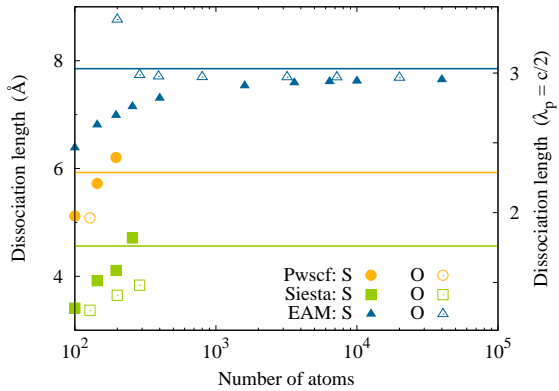}
  \caption{Dissociation length of the screw dislocation. 
  Symbols correspond to data extracted from atomistic simulations through the fit of the disregistry
  (Eq.~\ref{eq:PNperio}) for both the O and S periodic arrangements
  and for different sizes of the simulation cell.
  The solid lines are the predictions of elasticity theory based on the stacking faults
  (Tab.~\ref{tab:gFault}).
  On the right vertical axis, the experimental $c$ lattice parameter has been used 
  to normalized the dissociation length by the distance $\lambda_{\rm P}=c/2$ 
  between Peierls valleys.}
  \label{fig:dissociation}
\end{figure}

The dissociation length of the screw dislocation obtained through this fitting procedures
are shown in Fig.~\ref{fig:dissociation} for both periodic arrangements.
We observe variations with the dislocation periodic arrangement used in the simulation,
as well as with the size of the simulation cell. 
The results are nevertheless close to the predictions of 
elasticity theory (\S \ref{sec:dissociation}) for all the three interaction models. 
We could even see with the EAM potential that the dissociation lengths extracted from 
atomistic simulations converge to the value given by elasticity theory for large enough unit cells.
It is thus relevant to describe the screw dislocation as dissociated in two partial dislocations
linked by a stacking fault, although the dissociation length remains small.

\subsection{Core energy}

\begin{figure}[bth]
    \includegraphics[width=0.9\linewidth]{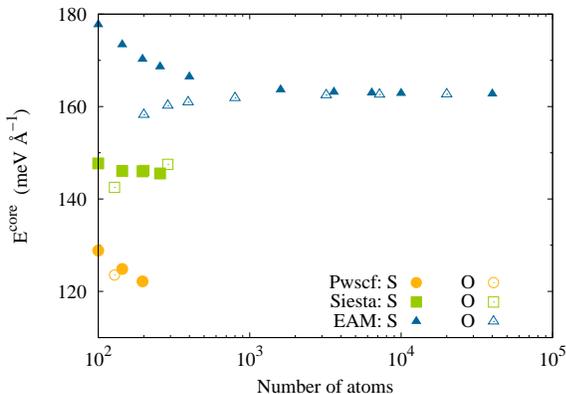}
  \caption{Core energy of the screw dislocation obtained for different
  sizes of the atomistic simulation cell and for both the O and S periodic arrangements
  ($r_{\rm c}=b$).}
  \label{fig:Ecore}
\end{figure}

We obtain the dislocation core energy by subtracting the elastic energy
from the excess energy given by the atomistic simulations.
This excess energy is the energy difference between the simulation cell
containing the dislocation dipole and the same cell without any defect.
The elastic energy calculation takes into account the elastic anisotropy\cite{Stroh1958,Stroh1962}
and the effect of periodic boundary conditions.\cite{Cai2003}
The obtained core energy are shown in Fig.~\ref{fig:Ecore}.
Like for the dissociation length, some variations with the size of the simulation cell
and the periodic arrangement can be observed.
We did not manage to link both quantities, \ie the core energy 
and the dissociation length.
They are not simply related by the expression of the energy variation 
with the dissociation length (Eq.~\ref{eq:Edissociation}), 
and the change of the elastic interaction between dislocations caused by their dissociation
could not fully explain the variation of the core energy either.
As the spreading of partial dislocations also depends on the size of the simulation cells, 
the variation of the core energy probably have a more complex origin 
than simply a variation of the dissociation length.

We could use, with the EAM potential, a simulation cell large enough
to obtain a converged value of the core energy, 163\,meV\,\AA$^{-1}$ 
for a core radius $r_{\rm c}=b$.
It is worth pointing that this value does not depend on the periodic
arrangement used in the simulation (Fig.~\ref{fig:Ecore}). This can be achieved thanks to a proper
account of the core traction contribution to the elastic energy.\cite{Clouet2009a}
A difference of $\sim10$\,meV\,\AA$^{-1}$ would have been observed 
between the S and O periodic arrangement without this contribution.
\Abinitio calculations lead to a dislocation core energy of 
$145\pm5$\,meV\,\AA$^{-1}$ for \siesta
and $125\pm5$\,meV\,\AA$^{-1}$ for \pwscf ($r_{\rm c}=b$ in both cases).

\subsection{Dissociation in the basal plane}

Basal slip is observed experimentally only at high temperature (above 850\,K)
and for a higher resolved shear stress than the one needed to activate prismatic slip.\cite{Akhtar1973}
At lower temperatures, no basal slip could be observed,\cite{Rapperport1960,Akhtar1971} 
even when the monocrystal was oriented so as to favor basal slip.
Our atomistic simulations lead to a screw dislocation configuration dissociated in the prism plane,
which clearly cannot glide easily in the basal plane. 
It is worth looking if another configuration, which could glide in this basal plane,
also exists.

\begin{figure}[bth]
    \includegraphics[width=0.8\linewidth]{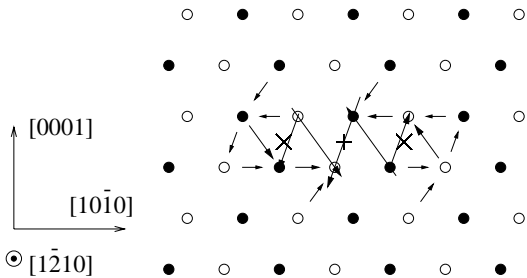}
  \caption{Differential displacement map of the metastable configuration 
  of a $1/3[1\bar{2}10]$ screw dislocation dissociated in the basal plane, 
  as obtained with the EAM potential for the O periodic arrangement with $n=6$ (288 atoms).
  Crosses $\times$ correspond to the positions of the two partial dislocations,
  and $+$ to their middle, \ie the position of the total dislocation.}
  \label{fig:vitek_basal}
\end{figure}

Using the same EAM potential, Khater and Bacon \cite{Khater2010} 
showed that a screw dislocation can also dissociate in the basal plane.
The basal configuration is obtained by introducing in the same basal plane
two partial dislocations of Burgers vector $1/3[1\bar{1}00]$ and $1/3[0\bar{1}10]$,
in agreement with the location of the minimum on the basal $\gamma$-surface (Fig.~\ref{fig:gSurfaceBasal}).
After relaxation of the atomic positions, the dislocation
remains dissociated in the basal plane, as can be seen from 
the corresponding differential displacement map (Fig.~\ref{fig:vitek_basal}).
A fit of the screw component of the disregistry created by the dislocation 
in the basal plane leads to a dissociation length $d=6.0$\,\AA, 
a higher value than predicted by elasticity theory (Tab.~\ref{tab:gFault}).
This configuration is metastable. It has indeed an energy higher by 62\,meV\,\AA$^{-1}$
than the configuration dissociated in the prism plane.

We check if \abinitio calculations also lead to such a metastable 
configuration. Starting from a screw dislocation dissociated 
in two partial dislocations in the basal plane, both \siesta
and \pwscf lead after relaxation of the atomic positions 
to the stable configuration dissociated in the prism plane. 
This is true both with the S ($n=5$) and the O ($n=4$) periodic arrangements. 
These \abinitio calculations show then that such 
a dissociation of the screw dislocation in the basal plane is unstable. 
The metastable configuration observed with the EAM potential
appears to be an artifact of the empirical potential.
This is not specific to the Mendelev and Ackland potential \cite{Mendelev2007}
as a configuration dissociated in the basal plane is stabilized 
by any central forces potential.\cite{Liang1986,Bacon2002,Khater2010}

\section{Peierls barrier}

Before calculating \abinitio the Peierls barrier of the screw dislocation, 
we use the EAM potential to assess the validity of the method 
and to check the convergence of the Peierls barrier with the size of the 
simulation cell.

\subsection{Methodology}

\begin{figure}[!bth]
    \includegraphics[width=0.9\linewidth]{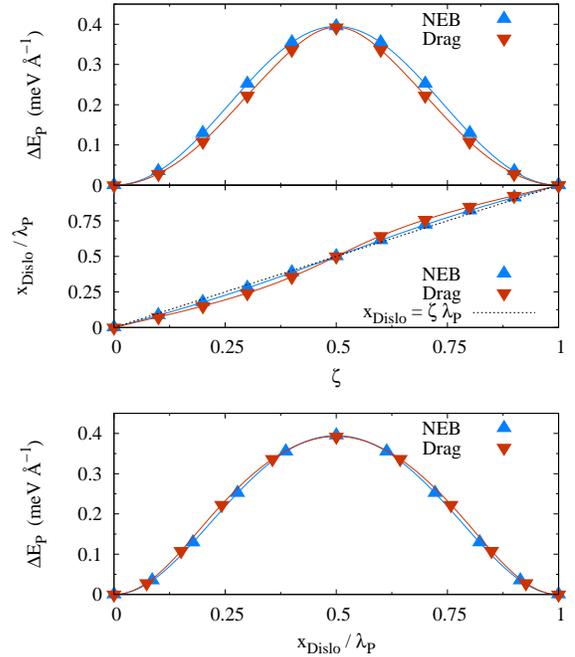}
  \caption{Peierls barrier of a screw dislocation calculated with the EAM potential
  for the S periodic arrangement with 1600 atoms ($n=m=20$). Two different methods 
  for finding the minimum energy path have been used: the NEB method 
  and a simple constrained minimization (drag).
  Symbols correspond to the results of atomistic simulations
  and lines to their interpolation with Fourier series.}
  \label{fig:Epeierls_neb_vs_drag}
\end{figure}

\begin{figure}[!bth]
    \includegraphics[width=0.9\linewidth]{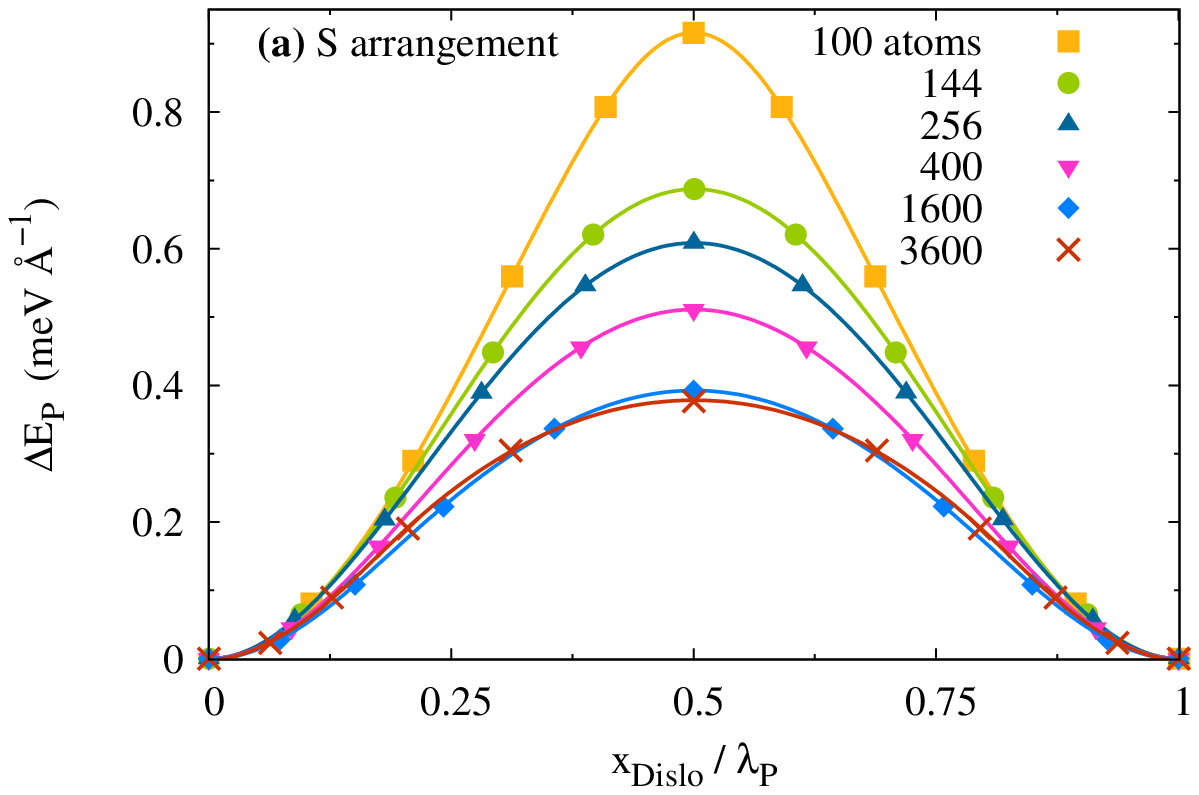}
    \includegraphics[width=0.9\linewidth]{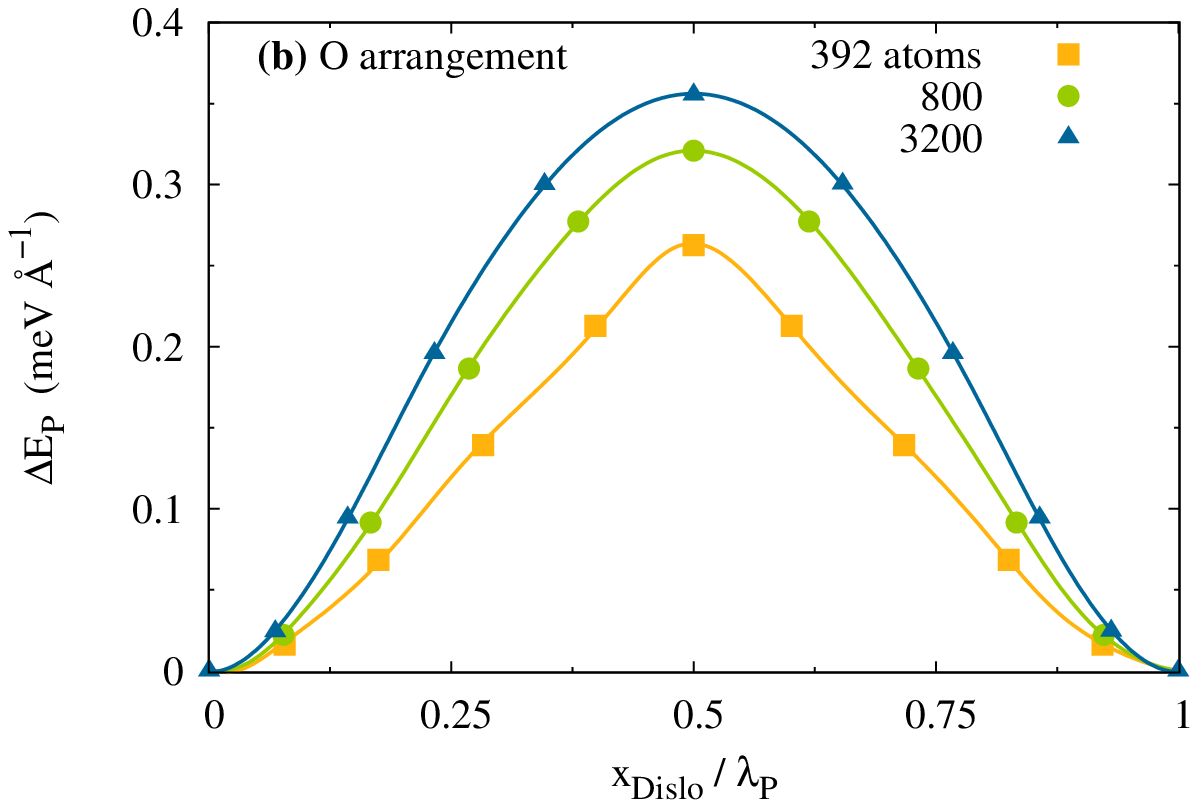}
  \caption{Variation of the Peierls barrier with the size of the simulation 
  cell calculated with the EAM potential for both periodic arrangements.}
  \label{fig:EpeierlsSize}
\end{figure}

We determine the Peierls barrier for the screw dislocation gliding
in the prism plane. This is done using a constrained minimization
between two adjacent equilibrium configurations of the dislocations.
We move both dislocations in the same direction
by one Peierls distance $\lambda_{\rm P} = c/2$ between the initial and final
states, so as to keep constant the distance between dislocations. 

Two different algorithms are used to perform the constrained minimization,
the simple drag method and the more robust nudged elastic band (NEB) method.
\cite{Henkelman2000a}
Intermediate configurations are built by linearly interpolating the atomic 
coordinates between the initial and final states.
We define the corresponding reaction coordinate 
$\zeta = (\vec{X}-\vec{X}^{\rm I}) \cdot (\vec{X}^{\rm F}-\vec{X}^{\rm I})
/ \| \vec{X}^{\rm F}-\vec{X}^{\rm I} \|^2$, where $\vec{X}$, $\vec{X}^{\rm I}$, 
and $\vec{X}^{\rm F}$ are the $3N$ vectors defining atomic positions
for respectively the intermediate, initial, and final configurations.
In the drag method, the minimization is performed on all atomic coordinates
with the constraint that $\zeta$ remains fixed for each of the nine intermediate
images. 
Nine intermediate images are also used in NEB method 
with a spring constant $k=0.1$\,eV\,\AA$^{-2}$.

So as to obtain the variation along the path of the dislocation energy 
with its position, we need to determine the dislocation position $x_{\rm Dislo}(\zeta)$
for each intermediate image, once it has been relaxed. 
This is done thanks to a fit of the disregistry in the prism plane, 
as described in \S\ref{sec:2:core}.
This allows us to check that both dislocations in the simulation cell 
are moving in a coordinated way: the distance between them remains fixed.
As a consequence, there is no variation of the elastic energy along the path
and the energy variation $\Delta E_{\rm P}(\zeta)$ obtained by the constrained minimization 
corresponds to a variation of the core energy, \ie the Peierls energy.
We deduce the Peierls energy $\Delta E_{\rm P}(x_{\rm Dislo})$ 
by eliminating the reaction coordinate $\zeta$ between $\Delta E_{\rm P}(\zeta)$
and $x_{\rm Dislo}(\zeta)$.

Fig.~\ref{fig:Epeierls_neb_vs_drag} illustrates the whole procedure 
for a given dislocation periodic arrangement using both constrained 
minimization techniques. 
The variation $\Delta E_{\rm P}(\zeta)$ slightly differs between both techniques:
for a given reaction coordinate $\zeta$, the drag method leads to a state
of lower energy than NEB method. 
This corresponds to a small difference in the dislocation position: 
this position deviates more with drag than with NEB method
from a linear variation.
Nevertheless, one obtains at the end the same Peierls barrier
$\Delta E_{\rm P}(x_{\rm Dislo})$, whatever the method used.
These differences observed for the functions $\Delta E_{\rm P}(\zeta)$
and $x_{\rm Dislo}(\zeta)$ between drag and NEB methods 
increase with the size of the simulation cell, \ie with the number
of degrees of liberty.
For large simulation cells (containing more than 3600 atoms 
in the S periodic arrangement for instance), the drag method 
sometimes fails to find a continuous path between the initial
and final states: one has to use the NEB method in theses cases.
Only much smaller simulation cells can be studied \abinitio. 
For these sizes, the drag and the NEB methods always lead to the same
result. We will therefore only use the drag method in the \abinitio
calculations, as it costs much less CPU time.

Finally, we interpolate with Fourier series the periodic functions
$\Delta E_{\rm P}(\zeta)$ and $x_{\rm Dislo}(\zeta)-\lambda_{\rm P} \zeta$. 
This leads to a smooth function $\Delta E_{\rm P}(x_{\rm Dislo})$ that can 
be derived. The Peierls stress $\sigma_{\rm P}$ is deduced from the 
maximal slope of this function, 
\begin{equation}
  \sigma_{\rm P} = \frac{1}{b}
  \Max{ \left(\frac{\partial \Delta E_{\rm P}}{\partial x_{\rm Dislo}}\right) }.
  \label{eq:sigmaP}
\end{equation}

We obtain a Peierls stress $\sigma_{\rm P} = 24 \pm 1$\,MPa
for the EAM potential. 
Khater and Bacon \cite{Khater2010} determined, for the same empirical potential, 
a Peierls stress $\sigma_{\rm P} = 22$\,MPa using molecular statics simulations
under applied stress. 
As the agreement between both methods is good, we see that the Peierls stress can be defined
either from the slope of the Peierls barrier 
or from the minimal applied stress under which the dislocation glides indefinitely.

We now examine, still with the EAM potential,
how this Peierls barrier varies when the size of the simulation cell decreases 
up to reaching a number of atoms that can be handled in \abinitio calculations 
(Fig.~\ref{fig:EpeierlsSize}). 
Both the S and O periodic arrangements give the same Peierls barrier, and hence the same 
Peierls stress, for a large enough simulation cell ($\gtrsim 1000$ atoms).
The Peierls barrier increases when the size of the simulation cell decreases with the S periodic arrangement, 
whereas it decreases with the O periodic arrangement. 
As a consequence, the S and O periodic arrangement should respectively lead
to an upper and lower limits of the Peierls stress for small simulation cells.
We will therefore use the S periodic arrangement to calculate \abinitio 
this Peierls barrier. This will allow us to confirm that the Peierls stress 
is as low as indicated by experiments and this EAM potential.
Moreover, it is worth pointing that the expected Peierls barrier should be small: 
$\Delta E_{\rm P}=0.4$\,meV\,\AA$^{-1}$ at the saddle point according to the EAM potential.
This corresponds to a difference of energy $2b\Delta E_{\rm P} = 2.6$\,meV
for a simulation cell of minimal height containing a dislocation dipole. 
Such a small energy difference may be problematic because of the precision 
of \abinitio calculations.
Looking for an upper limit of this value is easier as it minimizes the problems 
associated with this precision.

Finally, it is worth pointing that the dissociation length varies during the dislocation migration,
and this variation is more pronounced ($\sim10$\%) for the smallest simulation cells. 
But, like for the core energy, we did not manage to relate the size dependence of the Peierls barrier
to this variation of the dissociation length.

\subsection{\Abinitio barriers}

\begin{figure}[!bth]
    \includegraphics[width=0.9\linewidth]{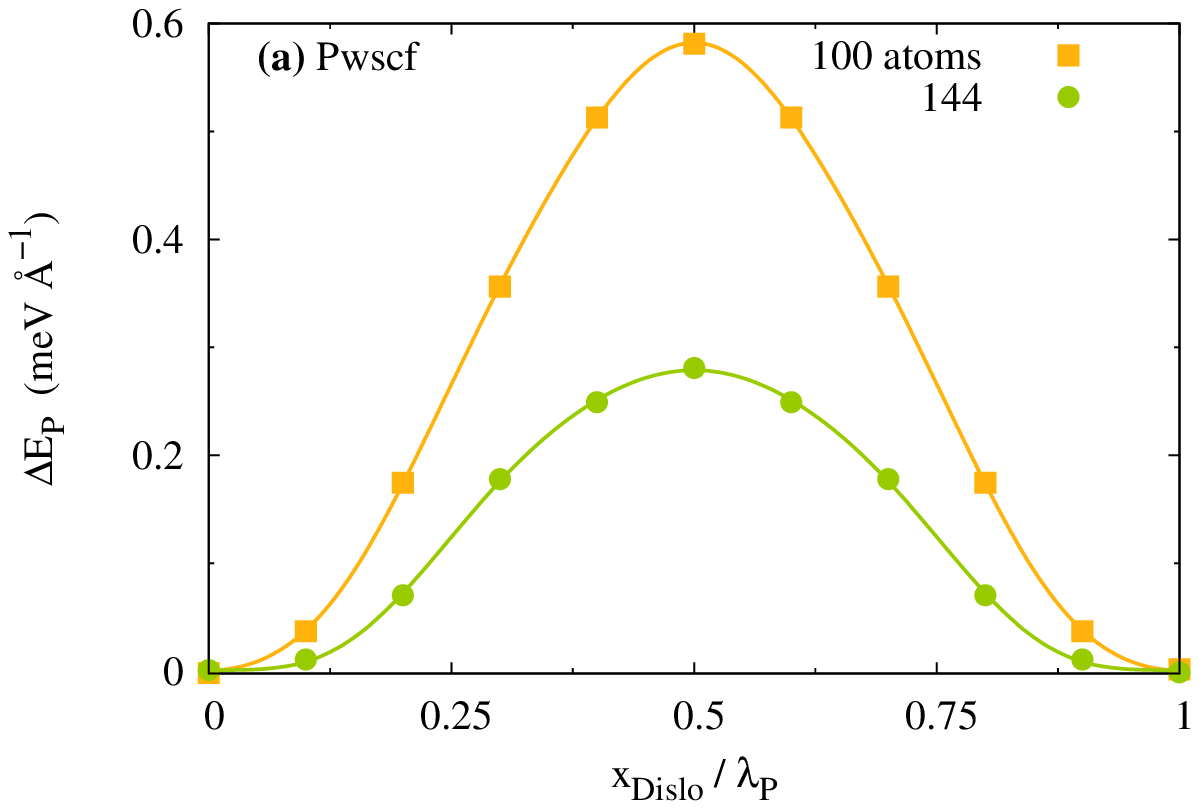}
    \includegraphics[width=0.9\linewidth]{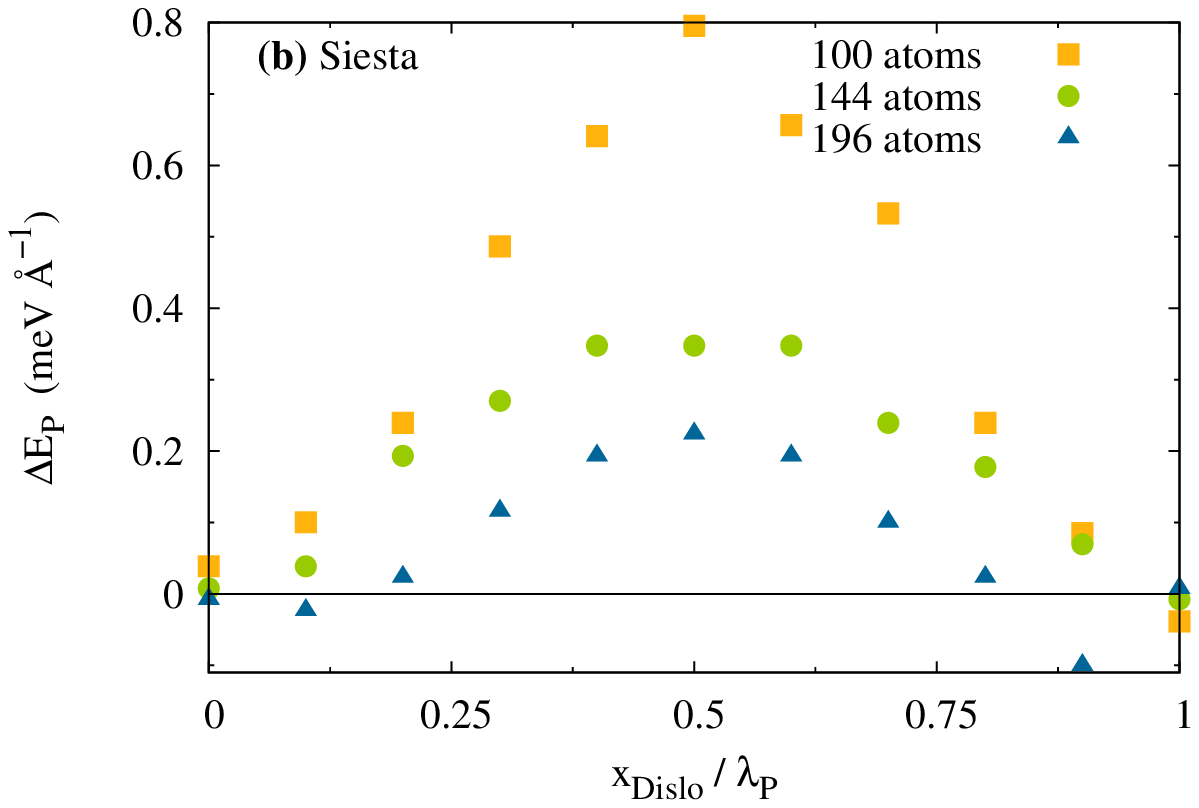}
  \caption{Peierls barrier for a screw dislocation gliding in its prism plane 
  calculated \abinitio with (a) \pwscf and (b) \siesta for the S periodic arrangement.
  Symbols correspond to \abinitio results and lines to their interpolation by Fourier series.}
  \label{fig:EpeierlsAbinitio}
\end{figure}

The Peierls barriers obtained by \abinitio calculations are shown 
on Fig.~\ref{fig:EpeierlsAbinitio}(a) for \pwscf
and Fig.~\ref{fig:EpeierlsAbinitio}(b) for \siesta. 
In both cases, the height of the barrier decreases when the number 
of atoms increases, in agreement with what has been observed with the EAM
potential for the same S periodic arrangement. 
For a given number of atoms, the \abinitio barriers are a little bit lower 
than the ones obtained with the EAM potential. 
Results obtained with \siesta are noisy: the energy barrier that we want to calculate is so small
that it needs a really strict convergence criterion on atomic forces for the relaxation.
We did not manage to reach such a precision with \siesta.  
On the other hand, the barriers obtained with \pwscf are smooth.
We can therefore interpolate the \abinitio results with Fourier series,
and estimate then the Peierls stress (Eq.~\ref{eq:sigmaP}). 
This leads to $\sigma_{\rm P}=36$\,MPa for the simulation cell containing 100 atoms
and $\sigma_{\rm P}=21$\,MPa for 144 atoms.
Considering that these values, obtained in small simulation cells,
are upper limits, 
\abinitio calculations predict a Peierls stress smaller than 21\,MPa
for a screw dislocation in zirconium gliding in a prism plane.

\begin{figure}[!bth]
    \includegraphics[width=0.9\linewidth]{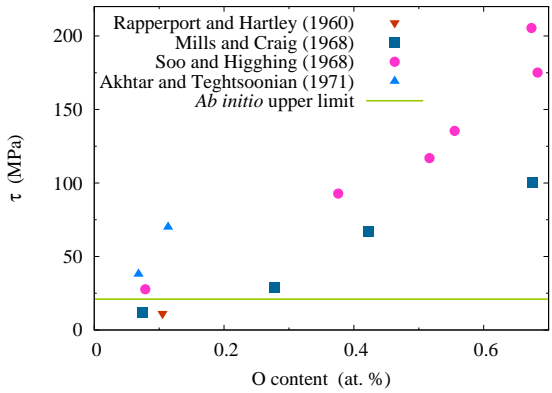}
  \caption{Zirconium flow stress determined experimentally for various O contents.
  \cite{Rapperport1960,Mills1968,Soo1968,Akhtar1971}
  Data have been extrapolated to 0\,K and are compared to the \abinitio Peierls stress
  of screw dislocations in pure Zr.}
  \label{fig:PeierlsOcontent}
\end{figure}

The comparison of this \abinitio estimate of the Peierls stress 
with experimental data\cite{Rapperport1960,Mills1968,Soo1968,Akhtar1971}
is quite challenging. 
As pointed out in the introduction, the yield stress of zirconium strongly depends
on its oxygen content. 
No experimental data exists for a purity better than 0.07\% O (in atomic fraction). 
Moreover, all measurements have been performed at temperatures higher than 77\,K.
The determination of pure zirconium flow stress at 0\,K therefore needs 
some extrapolation of experimental data. 
Without a clear understanding of the mechanisms responsible of zirconium hardening
by oxygen impurities, such an extrapolation is quite hazardous. 
Nevertheless, a graphical comparison (Fig.~\ref{fig:PeierlsOcontent})
of experimental data with our \abinitio Peierls stress 
shows a reasonable agreement, keeping in mind that the value 
21\,MPa has to be considered as an upper limit.

\section{Conclusions}

Using two \abinitio approaches (\pwscf and \siesta),
both in the DFT-GGA approximation,
we have shown that a $1/3\langle1\bar{2}10\rangle$ screw dislocation 
in hcp zirconium dissociates in two partial dislocations 
with a pure screw character.
This is in agreement with the minimum 
in $1/6\langle1\bar{2}10\rangle$ found for the generalized stacking fault
energy in the prism plane.
We could extract the dissociation length from our atomistic simulations. 
Although this dissociation length is small ($d\sim$6\,{\AA} for \pwscf
and $d\sim4$\,{\AA} for \siesta), it is in reasonable agreement with the 
one predicted by elasticity theory.

The EAM potential of Mendelev and Ackland\cite{Mendelev2007}
leads to the same structure of the dislocation dissociated 
in the prism plane. 
The metastable configuration dissociated in the basal plane 
predicted by this potential is not stable in \abinitio calculations,
both with \pwscf and \siesta. 
This configuration is therefore an artifact of this potential, 
probably induced by the deep minimum in $1/3\langle1\bar{1}00\rangle$
found with this empirical potential for the generalized stacking
fault in the basal plane.
This minimum is much more shallow in \abinitio calculations.

We could also obtain an \abinitio estimate of the Peierls stress
of the screw dislocation gliding in the prism plane. 
Calculations with \pwscf lead to an upper limit of 21\,MPa
for this Peierls stress.
This small value shows that screw dislocations can glide quite easily 
in pure zirconium, thus confirming what had been obtained with 
Mendelev and Ackland EAM potential. 
Such a small Peierls stress is in agreement with experimental data, 
once the hardening of oxygen impurities has been considered.

\begin{acknowledgments}
  This work was performed using HPC resources from GENCI-CINES and GENCI-CCRT
  (Grant Nos 2011-096020 and 2012-096847).
\end{acknowledgments}

\bibliographystyle{apsrev4-1}
\bibliography{clouet2012}

\end{document}